\documentclass[twocolumn,english,aps,prl,showpacs]{revtex4}
\usepackage{ulem,amsmath,graphicx,amssymb,epsfig,babel,dsfont}

\renewcommand{\Im}{{\rm Im}}
\newcommand{\Tr}{{\rm Tr}}
\newcommand{\rd}{{\rm d}}
\newcommand{\kb}{k_{\rm B}}

\newcommand{\ri}{{\rm i}}

\newcommand{\rs}{{\rm s}}
\newcommand{\rp}{{\rm p}}

\begin{document}

\title{On Super-Planckian thermal emission in far field regime}

\author{S.-A. Biehs}
\email{s.age.biehs@uni-oldenburg.de} 
\affiliation{Institut f\"{u}r Physik, Carl von Ossietzky Universit\"{a}t,
D-26111 Oldenburg, Germany.}

\author{P. Ben-Abdallah}
\email{pba@institutoptique.fr} 
\affiliation{Laboratoire Charles Fabry, Institut d'Optique Graduate School, CNRS,Universit\'{e} Paris-Saclay, 91127 Palaiseau, France. }

\date{\today}

\begin{abstract}
We study, in the framework of the Landauer theory, the thermal emission in far-field regime, of arbitrary indefinite planar media and finite size systems.  We prove that the flux radiated by the former is bounded by the blackbody emission while, for the second, there is in principle, no upper limit demonstrating so the possibility for a super-Planckian thermal emission with finite size systems. 
\end{abstract}

\maketitle

Since the pioneer works of Kirchoff~\cite{Kirchoff} and Planck~\cite{Planck} on the thermal emission radiated by a hot body, the blackbody was considered as the perfect thermal emitter.  Hence, it was admitted so far that no system could radiate more energy into the far-field than a blackbody at the same temperature. However, during the last decade, several  studies~\cite{Nefedov,Simovski,Sergeant} have claimed that some metamaterials can radiate energy beyond the blackbody limit allowing so a super-Planckian thermal emission. In this brief communication we investigate this problem using the Landauer formalism recently introduced to deal with radiative heat exchanges between 2 ~\cite{PBA2010,Biehs2010} or N objects~\cite{PBAEtAl2011,Messina,Riccardo,Riccardo2} both in near and far-field regimes. We first consider the problem of the upper bound for far-field thermal emission for arbitrary indefinite planar systems before focusing our attention on finite size systems.

To start, let us consider two arbitrary semi-infinite planar anisotropic media separated by a distance $d \gg \lambda_{\rm th}$ as sketched in Fig.~1. $\lambda_{\rm th}$ is the thermal wavelength given by Wien's law. According to the fluctuational electrodynamics theory~\cite{RytovBook1989} the radiative heat flux exchanged between these two media results from the thermal motion of microscopic charges within both materials which are held at a fixed temperature $T_{1}$ and $T_2$. The microscopic fluctuating charges lead to macroscopic fluctuating currents $\mathbf{J}^{\rm e}$ in each body which are the sources of fluctuating fields which can be formally written down as
\begin{align}
  \mathbf{E}(\mathbf{r},\omega) = \ri \omega \mu_0 \int_V\!\! \mathds{G}^{\rm EE}(\mathbf{r}, \mathbf{r}',\omega)\cdot\mathbf{J}^{\rm e}(\mathbf{r}''\omega), \\
   \mathbf{H}(\mathbf{r},\omega) = \ri \omega \mu_0 \int_V\!\! \mathds{G}^{\rm HE}(\mathbf{r}, \mathbf{r}',\omega)\cdot\mathbf{J}^{\rm e}(\mathbf{r}''\omega), 
\end{align}
where the integration is performed over the volume $V$ containing the source currents $\mathbf{J}^{\rm e}$; $\mu_0$ is the permeability of vacuum. 
The Greens functions $\mathds{G}^{\rm EE}$ and $\mathds{G}^{\rm HE}$ which establish the linear relations between the fields and the sources 
of the fields are connected by Faraday's law
\begin{equation}
   \mathds{G}^{\rm HE}(\mathbf{r}, \mathbf{r}',\omega) = \frac{1}{\ri \omega \mu_0} \nabla\times\mathds{G}^{\rm EE}(\mathbf{r}, \mathbf{r}',\omega).
\label{Eq:GEEGHE}
\end{equation} 
\begin{figure}[Hhbt]
\centering
\includegraphics[angle=0,scale=0.4]{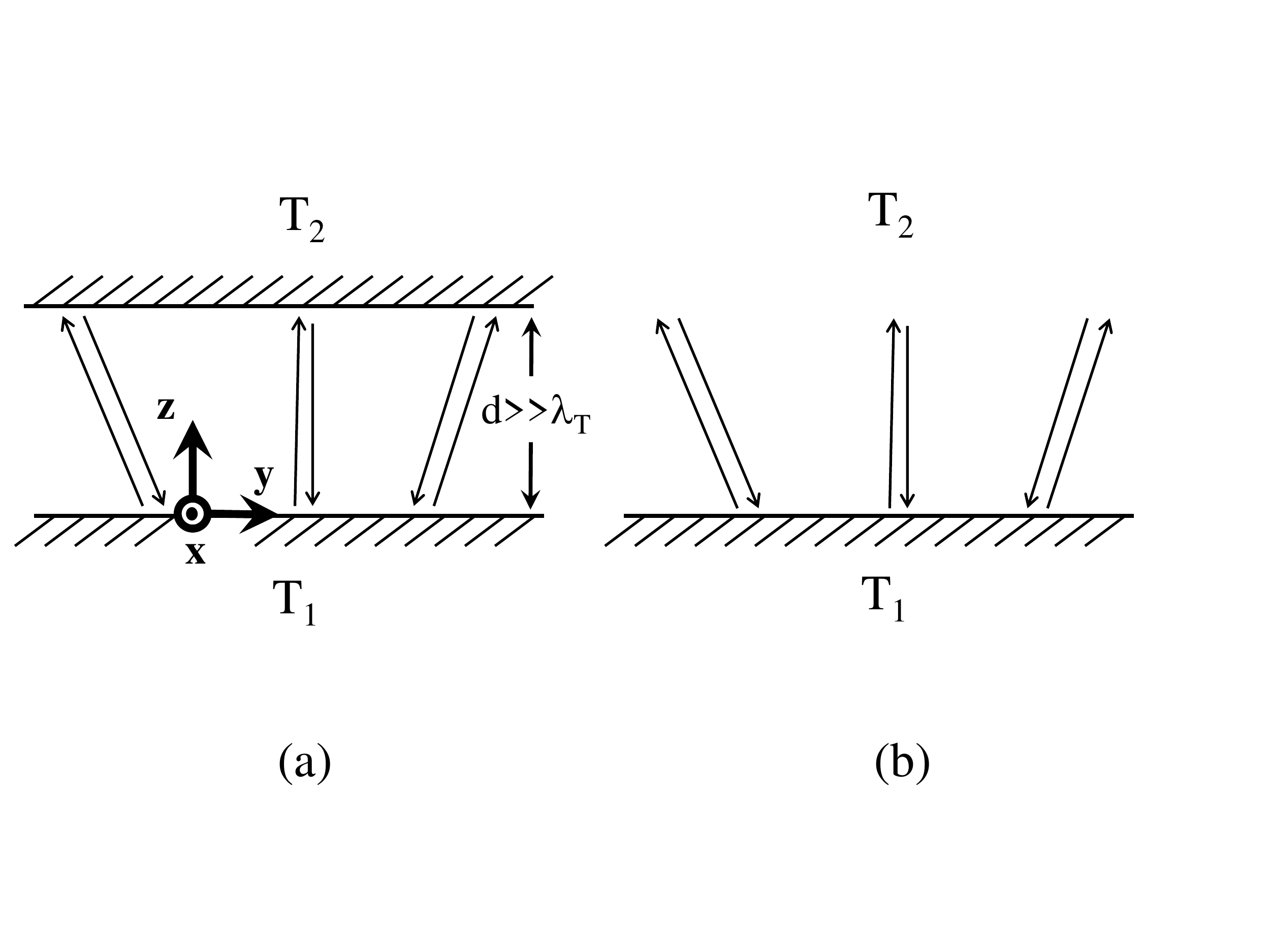}
\caption{Sketch of (a) two interacting arbitrary  planar systems and (b) a planar system in interaction with a thermal bath.}  
\end{figure} 
With the above expressions it is straight forward to determine the correlation functions of the fluctuating fields generated by the 
source currents. For our purpose we are interested in the correlation function 
$\langle E_\alpha (\mathbf{r},t)  H_\beta(\mathbf{r}',t') \rangle$ for $\alpha,\beta = x,y,z$ which are 
statistical averages of the fields with respect to ensembles of the fluctuating currents. Therefore it is necessary to know
the statistical properties of the source currents which are given according to the fluctuation-dissipation theorem by~\cite{Callen}
\begin{equation}
   \langle  J_\alpha^{\rm e}(\omega) J_\beta^{\rm e}(\omega') \rangle= 4 \pi \omega \Theta(T) \epsilon_0 \epsilon_{\alpha\beta}'' \delta(\mathbf{r-r'}) \delta(\omega + \omega'), 
\end{equation}
where $\epsilon_0$ is the permittivity of vacuum and $\uuline{\epsilon}''$ is the imaginary part of the permittivity 
tensor $\uuline{\epsilon}$ of the medium containing the source currents. The applicability of the fluctuation-dissipation theorem
requires that the media are at a local thermal equilibrium at temperature $T = T_1/T_2$. As a shorthand notation we have further 
introduced the mean energy of a harmonic oscillator at thermal equilibrium
\begin{equation}
  \Theta(T)=\frac{\hbar\omega}{2}+\frac{\hbar\omega}{e^{\hbar\omega \beta}- 1},
\end{equation}
which has in general contributions from vacuum and thermal fluctuations. Here $\beta = 1/(\kb T)$ is the inverse temperature
and $\kb$ is Boltzmann's constant. The part of the vacuum fluctuations can be neglected in the final expression since it does 
not contribute to the heat flux~\cite{RytovBook1989}. It follows that the correlation functions of field outside the media containing
the source currents are
\begin{equation}
\begin{split}
   \langle &E_\alpha (\mathbf{r},t)  H_\beta(\mathbf{r}',t') \rangle = \int_{-\infty}^{+\infty}\!\!\frac{d\omega}{2 \pi} e^{i \omega (t - t')} 2 \frac{\omega^3}{c^2} \mu_0 \Theta(T)\\
            &\quad \times \!\!   \int_V\!\! d^3 r'' \,\mathds{G}^{\rm EE}(\mathbf{r},\mathbf{r}'',\omega) \cdot \uuline{\epsilon}''(\mathbf{r}'') \cdot {\mathds{G}^{\rm HE}}^\dagger (\mathbf{r}',\mathbf{r}'',\omega),
\end{split}
\label{Eq:EHcorrelation}
\end{equation}
where $c$ is the vacuum light velocity. We emphasize that this expression is  general and is valid for any anisotropic non-magnetic material
with an arbitrary shape which is held at a fixed temperature $T$. Its generalization to magnetic materials is of
course straight forward.  From this expression we can determine the mean Poynting vector 
$\langle S_\alpha (\mathbf{r},t)\rangle=\eta_{\alpha\beta\gamma}\langle E_\beta (\mathbf{r},t)  H_\gamma(\mathbf{r},t) \rangle $ which determines
the amount of energy per unit time and unit area emitted by a medium at a given temperature.
Here $\eta_{\alpha\beta\gamma}$ denotes the total antisymetric Levi-Civita tensor.

In order to determine the heat radiated by a medium it is necessary to evaluate
expression~(\ref{Eq:EHcorrelation}) which can be done if the Green's function 
$\mathds{G}^{\rm EE}(\mathbf{r},\mathbf{r}'')$ is known. As for the magnetic Green's function, it
can then be calculated with Eq.~(\ref{Eq:GEEGHE}). That means we need the Green's function
with the source points $\mathbf{r}''$ inside the medium and the observation points $\mathbf{r}$
outside the medium. This procedure can be quite cumbersome in particular if the medium 
is anisotropic. In such cases it is useful to convert the volume integral into a surface
integral. Using Green's theorem we obtain
\begin{equation}
\begin{split}
\int_V\!\!d^3 \mathbf{r}' \mathds{G}^{\rm EE}(\mathbf{r},\mathbf{r}'',\omega)& \cdot \uuline{\epsilon}''(\mathbf{r}'') \cdot {\mathds{G}^{\rm HE}}^\dagger (\mathbf{r}',\mathbf{r}'',\omega) \\ &= - \frac{1}{2 \ri k_0^2}\mathds{I}^{\rm S} - \frac{1}{k_0^2} \Im \left(\mathds{G}^{\rm HE}(\mathbf{r,r}) \right) 
\end{split}
\end{equation} 
with the surface integral tensor 
\begin{equation}
\begin{split}
   \mathds{I}^{\rm S} &:= \int_{\partial V} \rd S' \biggl[ \bigl(\nabla'\times{\mathds{G}^{\rm EE}}^t(\mathbf{r},\mathbf{r}') \bigr)^t \cdot \bigl(\mathbf{n}\times{\mathds{G}^{\rm HE}}^\dagger (\mathbf{r,r'}) \bigr)\\ & \qquad + \mathds{G}^{\rm EE}(\mathbf{r},\mathbf{r}') \cdot\bigl(\mathbf{n}\times\nabla'\times{\mathds{G}^{\rm HE}}^\dagger(\mathbf{r,r'}) \bigr)  \biggr].
\end{split}
\label{Eq:SurfaceIntegralTensor}
\end{equation}
Here $\mathbf{n}$ is the surface normal on the boundary $\partial V$ of volume $V$; $t$ and $\dagger$ symbolize the transposition and
the hermitian conjugation of the Greens tensors and $k_0 = \omega/c$ is the wave vector in vacuum.
By means of this expression we can use Eq.~(\ref{Eq:EHcorrelation}) to write the mean Poynting vector as
\begin{equation}
\begin{split} 
   \langle S_\gamma (\mathbf{r},t) &\rangle=\eta_{\alpha\beta\gamma} \langle E_\alpha (\mathbf{r},t) H_\beta(\mathbf{r},t) \rangle \\
                                   &= \eta_{\alpha\beta\gamma} \int_0^\infty\!\!\frac{\rd \omega}{2 \pi} 2 \frac{\omega^3}{c^2} \mu_0 \Theta(T) \frac{\ri}{2 k_0^2} \mathds{I}^{\rm S}_{\alpha\beta} + \text{c.c.}
\end{split}
\label{Eq:PV}
\end{equation}
The advantage of this expression is obviously that it is only necessary to know the Greens function $\mathds{G}^{\rm EE}(\mathbf{r,r'})$ with
observation and source points outside the material. That means in particular that we do not need to determine the fields inside 
the medium itself. Furthermore we have replaced a volume integral by a surface integral which makes
the calculation simpler. Note that the same expression was found by Narayanaswamy and Zheng~\cite{Naraya}. 

In a planar geometry with a translational symmetry in x- and y-direction the Green tensor can be decomposed in plane waves. 
The resulting Weyl expression has the form
\begin{equation}
  \mathds{G}^{\rm EE}(\mathbf{r,r'}) = \int\!\!\frac{d^2 \kappa}{(2 \pi)^2} \mathds{G}^{\rm EE}(\mathbf{\kappa},z) e^{i \boldsymbol{\kappa} \cdot(\mathbf{x-x'})} \label{Eq:Green}.
\end{equation} 
The integral is a two-dimensional integral in $k_x$-$k_y$ space; $\boldsymbol{\kappa} := (k_x, k_y)^t$ and $\mathbf{x} =(x,y)^t$. 
For $z' < z$ the integrand $\mathds{G}^{\rm EE}(\mathbf{\kappa},z)$ can be written as~\cite{Opt_Exp}
\begin{equation}
\begin{split}
 \mathds{G}^{\rm EE}(\mathbf{\kappa},z) &=\frac{\ri}{2 k_z} \bigl[\mathds{D}_{12} \bigl( e^{\ri k_z (z - z')}\mathds{1}_{+}+ e^{\ri k_z (z + z')}\mathds{R}_1 \bigr)\\
                &\quad +\mathds{D}_{21} \bigl(\mathds{R}_2\mathds{R}_1 e^{-\ri k_z (z - z')} e^{2 \ri k_z d} \\
                &\quad+\mathds{R}_2e^{-\ri k_z (z + z')}e^{2\ri k_z d} \bigr) ]
\end{split}
\end{equation}
where $k_z = \sqrt{k_0^2 - \kappa^2}$ is the normal component of the wave vector. Here we have introduced the unit and reflection 
operators in polarization basis ($i,j = \rs,\rp$) 
\begin{align}
  \mathds{1}_{\pm}   &:= \sum_i \mathbf{a}_i^\pm \otimes \mathbf{a}_i^\pm, \\
  \mathds{R}_1 &:= \sum_{i,j} r_{ij}^{(1)} \mathbf{a}_i^+ \otimes \mathbf{a}_j^-, \\
  \mathds{R}_2 &:= \sum_{i,j} r_{ij}^{(2)} \mathbf{a}_i^- \otimes \mathbf{a}_j^+.
\end{align}
The polarization vectors for s- and p-polarization are defined as
\begin{equation}
  \mathbf{a}_\rs^\pm := \frac{1}{\kappa}\begin{pmatrix} k_y \\ -k_x \\ 0 \end{pmatrix}
\end{equation}
and
\begin{equation}
  \mathbf{a}_\rp^\pm := \frac{1}{\kappa k_0}\begin{pmatrix} \mp k_x k_z \\ \mp k_y k_z \\ \kappa^2 \end{pmatrix}.
\end{equation}
The reflection coefficients $r_{ij}^{(1/2)}$ are the Fresnel reflection coefficients of interface $1$ and $2$ 
describing how an incoming $j$-polarized wave is reflected into a $i$-polarized wave,
while $\mathds{D}_{ij}$ is the multiple scattering operator defined as
\begin{align}
    \mathds{D}_{12} &:= (\mathds{1}_{+} - \mathds{R}_1 \mathds{R}_2 e^{2 \ri k_{z} d})^{-1}, \\
    \mathds{D}_{21} &:= (\mathds{1}_{-} - \mathds{R}_2 \mathds{R}_1 e^{2 \ri k_{z} d})^{-1}.
\end{align}

It follows according to~(\ref{Eq:PV}) and (\ref{Eq:Green}) that the net flux $\Phi = \langle S_z \rangle$ (power per unit surface) exchanged between two arbitrary anisotropic media~\cite{Opt_Exp} separated by a distance $d$ larger than the thermal wavelength $\lambda_{\rm th}$ can be written into a Landauer-like form
\begin{equation}
    \Phi = 2\int_{0}^{\infty}\frac{\rd\omega}{2\pi}\,[\Theta(T_{1})-\Theta(T_{2})] \underset{\mid\boldsymbol{\kappa}\mid<\omega/c}{\int}\frac{\rd^2 \boldsymbol{\kappa}}{(2\pi)^2}  \mathcal{T}(\omega, \boldsymbol{\kappa}, d), \label{Eq:flux_inf}
\end{equation}
where we have defined the transmission coefficient $ \mathcal{T}$ as 
\begin{equation}
  \mathcal{T}(\omega, \boldsymbol{\kappa}, d) := \frac{1}{2}\Tr\bigl[(\mathds{1}_+-\mathds{R}_2^\dagger \mathds{R}_2)  \mathds{D}_{12}(\mathds{1}_+ - \mathds{R}_1 \mathds{R}_1^\dagger)  {\mathds{D}_{12}}^\dagger \bigr].
\label{Eq:TransmissionCoeff}
\end{equation}
Note that this expression is in accordance with results found by several other groups with different methods~\cite{Bimonte,Messina2011,Krueger2012}.
When the second medium is a bosonic field, then $\mathds{R}_2 =\mathds{O}$ (no reflecting medium) so that the transmission coefficient
simplifies to
\begin{equation}
   \mathcal{T}(\omega, \boldsymbol{\kappa}, d)=\frac{1}{2} \Tr\bigl[\mathds{1}_+ - \mathds{R}_1 \mathds{R}^\dagger_1] = 1 - \frac{1}{2}\parallel \mathds{R}_1\parallel_F^2,
\label{Eq:singleinterface}
\end{equation}
where
\begin{equation}
  \parallel \mathds{R}_1 \parallel_F^2 = \mid r^{ss}\mid^2+\mid r^{pp}\mid^2+\mid r^{sp}\mid^2+\mid r^{ps}\mid^2
\end{equation}
is the squared Frobenius norm of reflection operator. Since $ 2 \ge \parallel \mathds{R}_1 \parallel_F^2 \ge 0$, the net flux exchanged between the medium and its surrounding  is bounded by the maximal flux
\begin{equation}
 \Phi_{\rm max}=2\int_{0}^{\infty}\frac{\rd\omega}{2\pi}\,[\Theta(T_{1})-\Theta(T_{2})] \underset{\mid\boldsymbol{\kappa}\mid<\omega/c}{\int}\frac{\rd^2 \boldsymbol{\kappa}}{(2\pi)^2}.\label{Eq:flux_inf2}
\end{equation}
Since the $\kappa$-integral gives $\pi k_0^2$ (circle with radius $k_0$) 
\begin{equation}
\begin{split}
 \Phi_{\rm max}  &= \frac{c}{4} \int_0^\infty \!\! \rd \omega \bigl[\Theta(T_1) - \Theta(T_2) \bigr] \frac{\omega^2}{\pi^2 c^3} \\
                 &= \int_0^\infty \!\! \rd \omega \bigl[ I^0_\omega(T_1) - I^0_\omega(T_2) \bigr] \\
                 &= \sigma (T^4_1 - T^4_2)
\end{split}
\label{Eq:flux_inf3}
\end{equation}
where 
\begin{equation}
  \sigma := \frac{\pi^2 \kb^4}{60 c^2 \hbar^3} = 5.670373 \cdot 10^{-8}\:\rm W.\rm m^{-2}.\rm K^{-4}
\end{equation}
is Stefan-Boltzmann's constant and
\begin{equation}
    I^0_\omega (T) := \Theta(T) \frac{\omega^2}{4\pi^2 c^2}
\label{Eq:spectralBB}
\end{equation}
is the spectral intensity of a black body. The such derived upper bound (\ref{Eq:flux_inf3}) unambiguously proves that 
the power radiated by any planar isotropic or anisotropic material into its surrounding is always smaller or equal to the power that 
would be radiated by a blackbody at the same temperature. It is important to note that this limit exist not 
only for the total flux where the upper bound is set by the Stefan-Boltzmann's law but also spectrally where the
upper bound is set by $I^0_\omega(T_1) - I^0_\omega(T_2)$. 

The same limit applies of course also for the more general situation of radiative heat transfers between two
planar media. This is simply the case, because there can only be a maximal transmission into medium 2 if all
the incoming propagating waves are perfectly transmitted into medium 2. This is achieved if the reflectivity 
of medium 2 is zero, i.e.~$\mathds{R}_2 = \mathds{0}$. This is exactly the condition which lead to Stefan-Boltzmann's
law. Therefore the blackbody law provides the upper limit for heat radiation between planar materials even
if they are anisotropic. 

\begin{figure}[Hhbt]
\centering
\includegraphics[angle=0,scale=0.4]{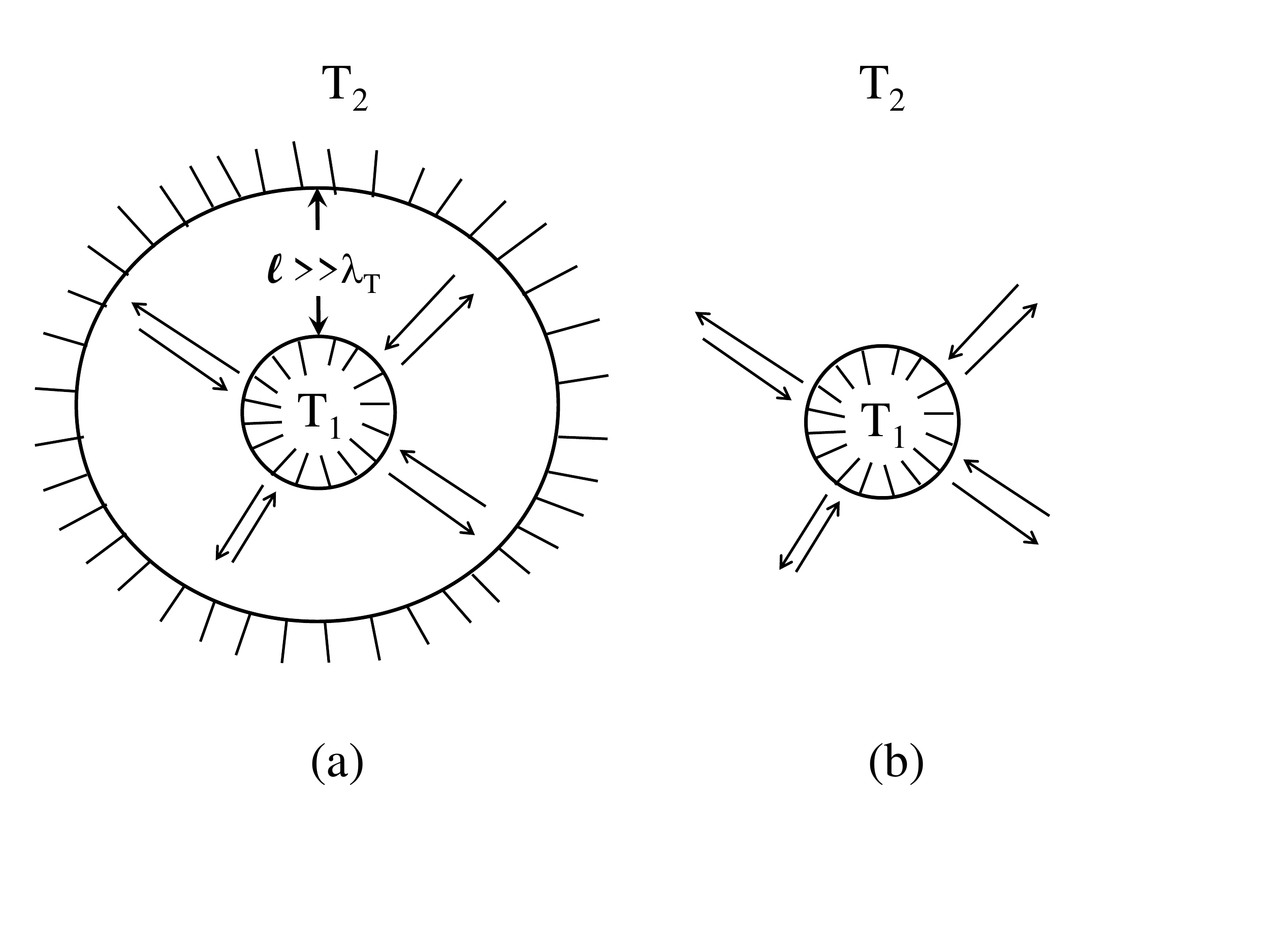}
\caption{Sketch of (a) a finite size medium in interaction with an encompassing system and (b) a finite size  system in interaction with a thermal bath.}
\end{figure}


Now, let us consider the case of finite size systems. A natural generalization of the previous configuration is the case of a sphere of radius $R$ encompassed by another sphere as illustrated in Fig.~2. As previously, the two bodies are hold at two different temperatures and separated by a shell of thickness $l \gg \lambda_{\rm th}$. By following the same course of action as in the plane-plane configuration, the net power exchanged between these media can be expressed in a Landauer-like form
\begin{equation}
  Q = \int_{0}^{\infty}\frac{\rd\omega}{2\pi}\,[\Theta(T_{1})-\Theta(T_{2})]  \mathcal{T}(\omega), \label{Eq:power_landauer}
\end{equation}
where the transmission coefficient $ \mathcal{T}(\omega)$ can be expressed in terms of the surface integral tensor
in Eq.~(\ref{Eq:SurfaceIntegralTensor}) (see Ref.~\cite{Naraya}).
Using $I^0$ from Eq.~(\ref{Eq:spectralBB}), the blackbody intensity at the frequency $\omega$, the net power exchanged between both media becomes
\begin{equation}
  Q = \int_{0}^{\infty}\rd\omega2\pi\frac{c^2}{\omega^2}[I^0_\omega(T_1)-I^0_\omega(T_2)]  \mathcal{T}(\omega). \label{Eq:power_Planck}
\end{equation}
Since $Q$ is the difference of the thermal emissions of the sphere and the wall it can also be written 
in terms of the thermal emissivity emissivity $\epsilon(\omega)$ which we introduce in the usual manner such that
\begin{equation}
  Q = A \int_{0}^{\infty}\rd\omega\epsilon(\omega)[I^0_\omega(T_1)-I^0_\omega(T_2)], \label{Eq:power_Planck2}
\end{equation}
where $A = 4\pi R^2$ is the surface area of the spherical body.
Hence, if the emissivity would be constant $\epsilon(\omega) \equiv \epsilon$, then we would obtain the
well-known expression 
\begin{equation}
   Q = A \epsilon \sigma (T_1^4 - T_2^4)
\label{Eq:BBsphere}
\end{equation}
for the exchanged power. The comparison of relations~(\ref{Eq:power_Planck}) and (\ref{Eq:power_Planck2}) shows 
that the relation between the emissivity and the transmission coefficient is
\begin{equation}
  \epsilon(\omega) = \frac{1}{2R^2}\frac{c^2}{\omega^2}\mathcal{T}(\omega). \label{Eq:emissivity_transmission}
\end{equation}
Of course, the emissivity is related to the absorptivity inside the spherical body~\cite{Bohren} --- as demanded by Kirchhoff's law ---
and the absorptivity itself can be written in terms of the absorption cross-section $\sigma_a(\omega)$. It follows
that~\cite{Bohren}
\begin{equation}
  \epsilon(\omega)=\frac{\sigma_a(\omega)}{\pi R^2} \label{Eq:emissivity_absorption}.
\end{equation}
Therefore, we have also a relation between the transmission coefficient and the absorption cross-section
\begin{equation}
  \mathcal{T}(\omega)=\frac{2}{\pi}\frac{\omega^2}{c^2}\sigma_a(\omega).\label{Eq:transmission_absorption}
\end{equation}
In arbitrary core-shell geometric configuration, the absorption cross-section reads~\cite{Qiu}
\begin{equation}
\begin{split}
   \sigma_a(\omega) &= \frac{\pi}{2}\frac{c^2}{\omega^2}\sum_{p={\rm TE,TM}} \sum_{l=1}^\infty (2l+1) (1-\mid r_{p,l}\mid^2),
\end{split}
\label{Eq:absorption_core-shell}
\end{equation}
where the summation is done over all channels (spherical waves) of TE and TM polarization, $r_{p,l}$ being the reflection coefficient of the system for the $l^{th}$ spherical order. It follows from relation~(\ref{Eq:transmission_absorption}) that the transmission coefficient takes the simple form
\begin{equation}
  \mathcal{T}(\omega) = \sum_{p={\rm TE,TM}} \sum_{l=1}^\infty (2l+1) (1-\mid r_{p,l}\mid^2).\label{Eq:transmission2}
\end{equation}
Contrary to the plane-plane configuration, a direct inspection of this serie shows that there is, in principle, no intrinsic upper 
limit for the flux $\Phi=\frac{Q}{A}$ (power per unit surface) exchanged between both media, where $A$ is the surface area of 
the inner sphere with radius $R$. Indeed, provided the medium can support higher order modes~\cite{Fan1, Fan2, JP_PBA} these modes
will increase the transmission coefficient such that the emissivity can become larger than one. Such behavior has been predicted 
long time ago with strong dissipating sphere in~\cite{Eisner}, for instance, and has also been discussed in several textbooks as in 
the famous Bohren and Huffman's book~\cite{Bohren}. Therefore Eq.~(\ref{Eq:BBsphere}) is not necessarily an upper limit in this case. 
And indeed, very recent works have found that core-shell particles~\cite{MaslovskiEtAl2016} and cylinders~\cite{GolykEtAl2012} can show 
a super-Planckian emission which is not a contradiction to the blackbody law, because it simply does not apply for finite 
size objects~\cite{Bohren}. 

In conclusion, we have shown that thermal radiation of a planar anisotropic medium is limited by Stefan-Boltzmann's law, so that
planar media cannot show Super-Planckian far-field emission. On the other hand, for finite size media Stefan-Boltzmann's does not
apply. As an example we discussed this for a spherical particle. In this case expression (\ref{Eq:transmission2}) for the transmission 
coefficient provides a natural target to be optimized within the Planck window in order to realize a finite size super-Planckian emitter. 
This optimization consists in minimizing the reflection coefficients for a maximum number of spherical channels and therefore to 
maximize the absorption cross-section of system. This is a direct consequence of reciprocity principle for the light as explicited 
by the generalized  Kirchoff law~\cite{RytovBook1989}.

\begin{acknowledgements}
\end{acknowledgements}

\end{document}